\documentstyle[preprint,aps]{revtex}
\begin{document}
\draft
\preprint{}
\title{Infinite Energy Dyon Solutions}
\author{D. Singleton}
\address{Department of Physics, Virginia Commonwealth University, 
1020 West Main St., Box 842000, Richmond, VA 23284-2000}
\date{\today}
\maketitle
\begin{abstract}
Three dyon solutions to the SU(2) Yang-Mills-Higgs system
are presented. These solutions are obtained from the BPS dyon
solutions by allowing the gauge fields to be complex or by
letting the free parameter of the BPS solution become
imaginary. In all cases however the physically measurable
quantities connected with these new solutions are entirely real. 
Although the new solutions are mathematically simple variations
of the BPS solution, they have one or more physically distinct
characteristics.
\end{abstract}
\newpage
\narrowtext

\section{The Dyon Solutions} 

The system studied in this paper is an SU(2) gauge theory
coupled to a scalar field in the triplet representation.
The scalar field is taken to have no mass or self interaction.
The Lagrangian for this system is
\begin{equation} 
\label{lagrange}
{\cal L} = -{ 1\over 4} G_{\mu \nu} ^a G^{\mu \nu} _a +
{1 \over 2} (D_{\mu} \phi _a ) (D^{\mu} \phi ^a )
\end{equation}
where $G_{\mu \nu} ^a$ is the field tensor of the SU(2)
gauge fields ($W_{\mu} ^a$) and $D_{\mu}$ is the covariant 
derivative of the scalar field. The equations of motion 
for this system are simplified through the use of a generalized
Wu-Yang ansatz \cite{yang} which was used by Witten \cite{wit}
to study multi-instanton solutions
\begin{eqnarray}
\label{wuyang}
W_i ^a &=& \epsilon _{aij} {r^j \over g r^2} [1 - K(r)] + 
\left( {r_i r_a \over r^2} - \delta _{ia} \right)
{G(r) \over g r} \nonumber \\
W_0 ^a &=& {r^a \over g r^2} J(r) \nonumber \\
\phi ^a &=& {r^a \over g r^2} H(r)
\end{eqnarray}
where $K(r)$, $G(r)$, $J(r)$, and $H(r)$ are the ansatz functions to be 
determined by the equations of motion. In terms of this ansatz the
field equations from the Lagrangian in Eq. (\ref{lagrange}) 
reduce to the following set of coupled, non-linear equations. 
\begin{eqnarray}
\label{difeq}
r^2 K'' &=& K (K^2 + G^2 + H^2 - J^2 - 1) \nonumber \\
r^2 G'' &=& G (K^2 + G^2 + H^2 - J^2 -1) \nonumber \\
r^2 J'' &=& 2J (K^2 + G^2) \nonumber \\
r^2 H'' &=& 2 H (K^2 + G^2)
\end{eqnarray}
where the primes denote differentiation with respect to $r$. The
solution to these equations discovered by Prasad and Sommerfield
\cite{prasad} and independently by Bogomolnyi \cite{bogo} is
\begin{eqnarray}
\label{soln}
K(r) &=& cos (\theta) C r \; csch(Cr) \nonumber \\
G(r) &=& sin (\theta) C r \; csch(Cr) \nonumber \\
J(r) &=& sinh(\gamma ) [1 - C r \; coth(C r)] \nonumber \\
H(r) &=& cosh (\gamma ) [ 1 - C r \; coth(C r)]
\end{eqnarray}
where $C$, $\theta$ and $\gamma$ are 
arbitrary constants. One of the nice
properties of this solution is that the energy in its fields is
finite (as compared to the field energy of a classical
point charged particle, which has a divergent energy from the
singularity at the origin). In terms of the ansatz functions
one can use standard Lagrangian techniques to obtain the energy
density of the fields
\begin{eqnarray}
\label{energy}
T^{00} &=& {1 \over g^2} \Bigg( {K'}^2 + {G'}^2 + 
{(K^2 + G^2 - 1)^2 \over 2 r^2} + {J^2 (K^2 + G^2) \over r^2} 
+ {(rJ' - J)^2 \over 2 r^2} \nonumber \\
&+& {H^2 (K^2 + G^2) \over r^2}
+ {(r H' - H)^2 \over 2 r^2} \Bigg)
\end{eqnarray}
For the solution in Eq. (\ref{soln}) this gives
\begin{equation}
\label{enbps1}
T^{00} = {2 cosh ^2 (\gamma) \over g^2} \left[ C^2 csch ^2 (C r)
\Big(1 - C r \; coth(C r) \Big)^2 + {\Big( C^2 r^2 csch ^2 
(C r) -1 \Big) ^2 \over 2 r^2} \right]
\end{equation}
This energy density can be integrated over all space to yield 
the total field energy of the BPS solution
\begin{equation}
\label{enbps2}
E = \int T^{00} d^3 x = {4 \pi C \over g^2} cosh ^2 (\gamma)
\end{equation}
To investigate the electromagnetic properties of this solution
't Hooft defined a generalized, gauge invariant, 
electromagnetic field strength tensor \cite{gthoof}
\begin{equation}
\label{emfst2}
F_{\mu \nu} = \partial _{\mu} (\hat{ \phi} ^a W^a _{\nu}) -
\partial _{\nu} (\hat{ \phi} ^a W^a _{\mu}) - {1 \over g}
\epsilon ^{abc} \hat{\phi } ^a ( \partial _{\mu} \hat{\phi } ^b )
( \partial _{\nu} \hat{\phi } ^c )
\end{equation}
where $\hat{\phi } ^a = \phi ^a (\phi ^b \phi ^b)^{-1/2}$. This
generalized U(1) field strength tensor reduces to the usual
expression for the field strength tensor if one performs a
gauge transformation to the ``Abelian'' gauge where the scalar
field only points in one direction in isospin space ({\it i.e.}
$\phi ^a = \delta ^{3a} v$) \cite{arafune}. Thus the electric
and magnetic fields of the BPS solution become 
\begin{eqnarray}
\label{ebfields}
{\cal E}_i &=& F_{i0} = {r_i \over g r} {d \over dr} \left( {J(r)
\over r} \right) = { sinh (\gamma) r_i \over g r^3}
(C^2 r^2 csch ^2 (C r) -1) \nonumber \\
{\cal B}_i &=& {1 \over 2} \epsilon _{ijk} F_{jk} = -{r_i \over
g r^3}
\end{eqnarray}
The magnetic field is that due to a point monopole of strength
$-4 \pi / g$ located at the origin. The electric field is that
due to some extended charge configuration of total charge 
$Q = - 4 \pi sinh (\gamma) / g$. The charge density which 
gives rise to the electric field can be caluclated  using
\begin{eqnarray}
\label{cdense}
\rho (r) &=& \nabla \cdot {\bf {\cal E}} ={1 \over r^2} {\partial
\over \partial r} (r^2 {\cal E}_r) \nonumber \\
&=& { 2 C^2 sinh (\gamma) csch ^2 (C r) \over g r} 
(1 - C r \; coth(C r))
\end{eqnarray}
Since the BPS solution has finite field energy,
this has led to its interpretation as a magnetically 
and electrically charged particle, whose mass is given by Eq.
(\ref{enbps2}). The ansatz functions $K(r)$, $G(r)$, $J(r)$, and
$H(r)$ (and therefore the gauge and scalar fields) are real.
If complex gauge fields and/or infinite energy configurations
are allowed there are several more solutions which can be found
for Eq. (\ref{difeq}). First by looking at the complementary
hyperbolic functions one finds the following complex
solution
\begin{eqnarray}
\label{soln1}
K(r) &=& {\bf i} cos(\theta ) C r \; sech(Cr) \nonumber \\
G(r) &=& {\bf i} sin(\theta ) C r \; sech(Cr) \nonumber \\
J(r) &=& sinh(\gamma ) [1 - C r \; tanh(C r)] \nonumber \\
H(r) &=& cosh (\gamma ) [ 1 - C r \; tanh(C r)]
\end{eqnarray}
Since the ansatz functions $K(r)$ and $G(r)$ are now imaginary, the 
space components of the gauge fields will be complex. Despite this one
finds that all the above listed quantities ({\it i.e} the field 
energy density, the electric field, the magnetic field, and 
the charge density) associated with this complex solution are
completely real. Although this solution was found by making some
simple changes we will see that it has some physical features which are
distinct from the BPS solution ({\it e.g.} the electric charge density
and the energy density). Inserting the ansatz functions of Eq. 
(\ref{soln1}) into Eq. (\ref{energy}) we find that the field 
energy density is
\begin{equation}
\label{ensoln1}
T^{00} = {2 cosh ^2 (\gamma) \over g^2} \left[ - C^2 sech ^2 (C r)
\Big( 1 - C r \; tanh(C r) \Big) ^2 + {(C^2 r^2 sech ^2 (C r) + 1)^2
\over 2 r^2} \right]
\end{equation}
The energy density is real even though the space components of the
gauge fields are complex. However the total energy in this field
configuration is infinite due to the singularity in the energy
density at $r = 0$. Thus the above solution is more like a Wu-Yang
monopole \cite{yang} or a charged point particle, as opposed
to a BPS dyon which has non-singular fields and finite energy. Using 
Eq. (\ref{ebfields}) we find that the electric and magnetic fields
associated with the solution in Eq. (\ref{soln1}) are
\begin{eqnarray}
\label{eb1}
{\cal E}_i &=& { - sinh (\gamma) r_i \over g r^3}
(C^2 r^2 sech ^2 (C r) + 1) \nonumber \\
{\cal B}_i &=& -{r_i \over g r^3}
\end{eqnarray}
The complex solution has the same $- 4 \pi / g$ magnetic charge
as the BPS solution. The electric field is that of some finite 
distribution of electric charge of total charge $ - 4 \pi sinh 
\gamma / g$. This is the same as the total electric charge 
carried by the BPS solution. By using Eq. (\ref{cdense}) 
we find that the electric charge density for the complex 
solution is given by
\begin{equation}
\label{cd1}
\rho (r) ={ - 2 C^2 sinh (\gamma) sech ^2 (C r) 
\over g r} \Big( 1 - C r \; tanh(C r) \Big)
\end{equation}
This charge density is real, has a singularity at the origin, and
falls off exponetially for large $r$.
Even though the space components of the gauge fields are complex
all the physical quantities calculated from it are real. The main
difference between this solution and the BPS solution is the
infinite field energy of the complex solution.

To obtain the next solution we apply the transformation, $C
\rightarrow i C$ to the BPS solution of Eq. (\ref{soln}). This
changes the hyperbolic functions into their trigonometric
counterparts, and yields the following solution to Eq. (\ref{difeq})
\begin{eqnarray}
\label{soln2}
K(r) &=& cos( \theta ) C r \; csc(Cr) \nonumber \\
G(r) &=& sin ( \theta ) C r \; csc(Cr) \nonumber \\
J(r) &=& sinh(\gamma ) [1 - C r \; cot(C r)] \nonumber \\
H(r) &=& cosh (\gamma ) [ 1 - C r \; cot(C r)]
\end{eqnarray}
This solution is completely real as opposed to the hyperbolic 
solution in Eq. (\ref{soln1}). Even though this solution
was obtained from the original BPS solution via a trivial
transformation, it has very different physical properties.
Most obviously the ansatz functions, and therefore the gauge
and scalar fields, become singular when $C r = n \pi$
where $n = 1, 2, 3, 4 ... \;$ . Thus this solution exhibits a series
of concentric spherical surfaces on which its fields become
singular. These singularities also show up in the energy density
of this solution. Inserting the ansatz
functions of Eq. (\ref{soln2}) in Eq. (\ref{energy}) we find
that the energy density of this solution is
\begin{equation}
\label{ensoln2}
T^{00} = {2 cosh ^2 (\gamma) \over g^2} \left[ C^2 csc ^2 (C r)
\Big( 1 - C r \; cot(C r)\Big) ^2 + {\Big( C^2 r^2 
csc ^2 (C r) - 1 \Big) ^2 \over 2 r^2} \right]
\end{equation}
The energy density becomes singular on the same spherical surfaces 
as the gauge and scalar fields. These spherical shells on which
the energy density becomes infinite cause the total field
energy of this solution $(E = \int T^{00} d^3 x)$ to diverge.
The electric and magnetic fields of this solution are obtained
using Eq.(\ref{ebfields})
\begin{eqnarray}
\label{eb2}
{\cal E}_i &=& { - sinh (\gamma) r_i \over g r^3}
(C^2 r^2 csc ^2 (C r) - 1) \nonumber \\
{\cal B}_i &=& -{r_i \over g r^3}
\end{eqnarray}
The magnetic field is exactly the same as that of the BPS solution
or the solution of Eq. (\ref{soln1}). The electric field is very
unusual for this solution. First, because of the $C^2 r^2 csc ^2 (C r)$
term, the electric field does not fall off for large $r$, but has
a periodic behaviour. Second, the electric field becomes
singular on the spherical shells given by $C r = n \pi$. One could
take the electric charge of this solution as located on these
singular surfaces. Finally, the electric charge of this solution 
is infinite, as is indicated by the electric field or by directly
looking at the charge density
\begin{equation}
\label{cd2}
\rho (r) ={ 2 C^2 sinh (\gamma) csc ^2 (C r) 
\over g r} \Big(1 - C r \; cot (C r) \Big)
\end{equation}
The infinite electric charge of this solution is its worst
feature. However, for the special case of this solution
where $\gamma = 0$ , one finds that the solution carries
no electric charge, but only a magnetic charge. Even in this
case the energy density still becomes singular on the
concentric spherical surfaces. 
Though this solution was obtained from the original BPS
solution by a trivial transformation, its physical characteristics
are different. Both the BPS solution and the solution from
Eq. (\ref{soln1}) had finite magnetic and electric charges
for any finite value of $\gamma$.
Thus both solutions could be viewed as dyonic particles (The BPS 
solution had the additional nice feature that its total
field energy was finite.). The solution given in Eq. (\ref{soln2}),
while having the same magnetic charge as the other two solutions, 
has an infinite electric charge in the general case when 
$\gamma \ne 0$. Although this solution is a
dyon in the sense that it carries both magnetic and electric charge
it is probably not correct to view it as a particle-like
solution. At this point
it is unclear how one should view this solution. The spherical
singular surfaces of this solution are similiar to that of
the Schwarzschild-like solution presented in 
Ref. \cite{sing}. However, the solution in Ref.
\cite{sing} only possessed one spherical surface on which the
fields and energy density diverged, and it carried no net electric
charge. When the solution of Ref. \cite{sing} was treated as a
background field in which a test particle was placed it was
found that the spherical singularity acted as an impenetrable
barrier which would trap the test particle either in the interior
or the exterior of the sphere \cite{yoshida}. Similiar results have
been found for other singular solutions \cite{swank} \cite{lunev}.
Treating the present solution as a background field would trap the
test particle between any two of the concentric spherical singularities.

Finally one can obtain a third solution to Eq. (\ref{difeq}) by
applying the transformation $C \rightarrow i C$ to the complex
solution in Eq. (\ref{soln1}). This yields
\begin{eqnarray}
\label{soln3}
K(r) &=& - cos( \theta ) C r \; sec(Cr) \nonumber \\
G(r) &=& - sin( \theta ) C r \; sec(Cr) \nonumber \\
J(r) &=& sinh(\gamma ) [1 + C r \; tan(C r)] \nonumber \\
H(r) &=& cosh (\gamma ) [ 1 + C r \; tan(C r)]
\end{eqnarray}
This solution is basically the same as the one obtained by
taking $C \rightarrow i C$ for the BPS solution. Most of
the comments concerning the solution in Eq. (\ref{soln2})
apply here as well. The singularities in the fields and energy
density are now located at $r =0$ and on the spherical surfaces
$C r = n \pi / 2$ where $n =1,3,5,7 ...$. The solution also has
infinite total field energy, and infinite electric charge,
unless $\gamma = 0$. As with all the other solutions it 
possesses a magnetic charge of $ - 4 \pi / g$.

Many of the physical characteristics of the solutions were
substanstially different in each case. However the magnetic
field and magnetic charge of all the solutions were the same.
This comes about since the magnetic charge of each solution is
a topological charge which carries the same value for each
field configuraion. To investigate this we look at the
topological current, $k _{\mu}$, of  Ref. \cite{arafune}
\begin{equation}
\label{current}
k_{\mu} = {1 \over 8 \pi} \epsilon_{\mu \nu \alpha \beta} 
\epsilon _{abc} \partial ^{\nu} {\hat \phi} ^a \partial ^{\alpha} 
{\hat \phi} ^b \partial ^{\beta} {\hat \phi} ^c
\end{equation}
The topological charge of this field configuration is then
\begin{eqnarray}
\label{tc}
q &=& \int k_0 d^3 x  = {1 \over 8 \pi} \int 
(\epsilon_{ijk} \epsilon_{abc}
\partial ^{i} {\hat \phi} ^a \partial ^{j} {\hat \phi} ^b
\partial ^{k} {\hat \phi} ^c ) d^3 x \nonumber \\
&=& {1 \over 8 \pi} \int \epsilon_{ijk} \epsilon_{abc}
\partial ^{i} ( {\hat \phi} ^a \partial ^{j} {\hat \phi} ^b
\partial ^{k} {\hat \phi} ^c ) d^3 x
\end{eqnarray}
For all the solutions one finds that ${\hat \phi } ^a = r^a / r$ 
which is the same regardless of the ansatz function $H(r)$. In all
cases we find that the topological charge is $q =1$. Since the
magnetic charge is identified with this topological charge via
't Hooft's generalized electromagnetic tensor, the magnetic charge
of each solution is the same.

\section{Discussion and Conclusion}

In this paper we have presented three new, exact solutions for
the SU(2) Yang-Mills-Higgs system, where the scalar field has
vanishing mass and self coupling. All three solutions were
connected to the well known BPS solution either by allowing the 
gauge fields to be complex or by allowing the free parameter in the 
BPS solution to be imaginary. In the first solution given in
Eq. (\ref{soln1}) we replaced the hyperbolic functions of the BPS
solution with their complements and let the ansatz 
functions, $K(r)$ and $G(r)$,
be imaginary, thus making the space components of the gauge fields
complex. Despite having complex gauge fields, this solution 
gave real results for all the physical quantities
calculated from it ({\it i.e.} the field energy density,
the electric and magentic fields, and the topological index of
the solution were all real). The magnetic and electric fields of this
solution indicated that it was a dyon carrying a magnetic charge
of $-4 \pi / g$ and and electric charge of $- 4 \pi sinh(\gamma ) /
g$, with the magnetic charge being concentrated at the origin while
the electric charge was spread out in an exponentially decaying
distribution around the origin. The only drawback of this solution 
is that, unlike the BPS solution, it has a divergent field energy 
due to the singularity in the field energy density 
at the origin. Thus, although
this solution looks mathematically similiar to the BPS solution, 
physically it may be more correct to think of it as a dyonic
version of the Wu-Yang monopole, which also has divergent 
field energy due to a singularity at $r=0$. 

The two other new solutions presented here, were derived from the BPS
solution and the complementary solution of Eq. (\ref{soln1}) by
allowing the free parameter $C$ of these solutions
to become imaginary ({\it i.e.}
$C \rightarrow i C$). This resulted in solutions with completely
real gauge and scalar fields, and with the hyperbolic functions
replaced by their trigonometric counterparts. In both cases the 
solutions carried the same magnetic charge as the previous hyperbolic
solutions. Now however the gauge and scalar fields developed singularities
on an infinite series of concentric spherical shells (the solution
in Eq. (\ref{soln3}) also had a singularity at $r=0$). In addition, 
unless $\gamma = 0$, both solutions carry an infinite 
electric charge, which can be thought of as concentrated on the 
spherical shells. Thus, although in the general case these solutions 
do carry both magnetic and electric charge, it is difficult to see
how they could be viewed as particle-like, and the name dyon is probably
inappropriate (even in the case of the complex solution of Eq. 
(\ref{soln1}), which had infinite field energy, the fact that it
had localized charges, and a localized divergent energy density
makes a particle-like intrepretation reasonable).
The spherically, singular surfaces of
the trigonometric solutions are somewhat similiar to the spherically
singular surface of the Schwarzschild-like solution of Ref. 
\cite{sing}. The Schwarzschild-like solution, however, had only one
spherically singular surface rather than an infinite series of
concentric surfaces, and it carried no electric charge rather than
the infinite electric charge carried by the trigonometric
solutions in the general $\gamma \ne 0$ case. 
At present it is not clear what intrepretation 
can be given to these trigonometric solutions or
what physcial role if any they may play. One could argue that
the singularities in the trigonometric solutions and to a lesser
extent the singularity in the solution of Eq. (\ref{soln1}) might
indicate that they are not physically interesting. However  this
is not necessarily the case as can be seen by the example of the meron
solution \cite{dea} which is singular, and yet is thought to play 
a role in some non-Abelian gauge theories \cite{callan}. One might 
consider using the trigonometric solutions to solve the field
equations in some finite region around the origin, and patching 
them together with one of the other solutions (either the
BPS solution or the Schwarzschild-like solution) which
are better behaved as $r \rightarrow \infty$. This would put some
conditions on the fields and their derivatives at the point where
the solutions are joined, thus possibly fixing some of the
arbitrary constants ({\it e.g.} $C$) of the solutions. In one sense
the trigonometric solutions, despite their singularities, are
interesting solutions since they have entirely different 
physical characteristics from any previously given 
solutions to the SU(2) Yang-Mills-Higgs system.

\section{Acknowledgements} The author would like to thank 
Tina Ilvonen and Susan Davis for  help and encouragement.

\end{document}